\documentclass[12pt,a4paper]{article}

\newcommand{\expect}{\mathsf{E}}
\newcommand{\var}{\mathsf{var}}
\newcommand{\cov}{\mathsf{cov}}
\newcommand{\diag}{\mathrm{diag}}
\newcommand{\iv}{\sigma^2_e}

\newcommand{\R}{$\mathsf{R\;}$}
\usepackage{color}
\usepackage{amsfonts}
\usepackage{amssymb,amsmath}
\usepackage{mathtools}
\usepackage{mathrsfs}
\usepackage{graphicx}
\usepackage{bm}
\usepackage{natbib}
\usepackage{wasysym}
\usepackage[affil-it]{authblk}

\usepackage{pdflscape}

\begin{document}

\title{Highly Efficient Stepped Wedge Designs for Clusters of Unequal Size}

\author
{J.N.S. Matthews
\thanks{\texttt{john.matthews@ncl.ac.uk}: corresponding author}}

\affil{School of Mathematics, Statistics \& Physics and Institute of Health \& Society, \\ Newcastle University,\\ Newcastle upon Tyne, NE1 7RU, U.K.\\}

\date{}

\maketitle

\begin{abstract}
The Stepped Wedge Design (SWD) is a form of cluster randomized trial, usually comparing two treatments, which is divided into time periods and sequences, with clusters allocated to sequences.  Typically all sequences start with the standard treatment and end with the new treatment, with the change happening at different times in the different sequences.  The clusters will usually differ in size but this is overlooked in much of the existing literature. This paper considers the case when clusters have different sizes and determines how efficient designs can be found.  The approach uses an approximation to the the variance of the treatment effect which is expressed in terms of the proportions of clusters and of individuals allocated to each sequence of the design.  The roles of these sets of proportions in determining an efficient design are discussed and illustrated using two SWDs, one in the treatment of sexually transmitted diseases and one in renal replacement therapy. Cluster-balanced designs, which allocate equal numbers of clusters to each sequence, are shown to have excellent statistical and practical properties;  suggestions are made about the practical application of the results for these designs. The paper concentrates on the cross-sectional  case, where subjects are measured once, but it is briefly indicated how the methods can be extended to the closed-cohort design.
\end{abstract}
Key words {\it
Closed-cohort design; Cluster randomized trial; Cross-sectional design; Optimal design; Stepped wedge design.}
\maketitle
\section{Introduction}
Recently there has been interest in the Stepped Wedge Design (SWD), a type of cluster-crossover design which is divided into $T$ treatment periods and comprises $S$ treatment sequences, and usually compares two treatments, A and B. The main characteristic of a SWD is that each sequence allows only one change in treatment, from A to B say, and reversion to A is not permitted.  In the classic SWD shown in Table \ref{tab1}, all sequences start on treatment A and finish on treatment B: clusters are allocated to sequences and the treatment administered changes from A to B in a period determined by the sequence.  Typically A is the standard treatment and B is a new treatment, so SWDs are particularly well suited to implementation studies, where the staggered change from standard to new can have practical advantages: see \citet{mdege} for an early review and \citet{Barker} for issues related to statistical methodology.  Although there may be circumstances where they offer substantial practical benefits, these must be balanced against some statistical weaknesses that are inherent in the design \citep[see][]{matthewsforbes}.\\
\begin{table}[!h]
\begin{center}
\caption{A stepped wedge design with $S$ sequences, comparing treatment A and B over $T$ periods, $S=T-1$.}
\label{tab1}
\begin{tabular}{c|cccccc}
\hline
&\multicolumn{6}{c}{Time}\\
Sequence & 1& 2& 3&$\cdots$& $T-1$ & $T$\\
\hline
1& A & A& A& $\cdots$& A& B \\
2& A & A& A& $\cdots$ &B& B \\
$\vdots$ &&&&$\cdots$&\\
$S-1$&A&A&B&$\cdots$ & B & B\\
$S$ &A&B&B&$\cdots$ & B& B\\
\end{tabular}
\end{center}
\end{table}
The analysis of continuous outcomes is often based on the linear mixed effects model proposed by \citet{HusseyHughes}.  Issues regarding the design of SWDs are frequently discussed in terms of this model and include: consideration of samples size calculations \citep{HusseyHughes,Hemmingtaljaard1}; choosing the number of sequences and length of periods \citep{CopasTrialsdes}; and modifications of the classic design, \citep[e.g.][]{JennT1,HemmGirling2}. It should be noted that the model in \citet{HusseyHughes} applies to \emph{cross-sectional} SWDs, where within a cluster different individuals are observed in different periods: the model would need modification for a \emph{closed-cohort} design, where individuals in a cluster are observed in every period of the study.  \\
The principal aim of a SWD is to estimate the treatment effect, $\theta$.  As the information on $\theta$ accrues differently on the different sequences \citep{kaszaforbes}, how the clusters are allocated to the sequences can affect $\var(\widehat\theta)$, which means that some allocations may be more efficient than others.  Most work has assumed that clusters are of the same size and that equal numbers of clusters are allocated to each of the $S$ sequences.  For cross-sectional studies with clusters of equal size \citet{forbesopt} showed that allocating a larger proportion of clusters to sequences 1 and $S$ than to the other sequences gave a more efficient design, with the degree of imbalance between the proportions depending on the intra-class correlation (ICC), $\rho$: the extension to the closed-cohort case was addressed in \citet{Lietal}.\\
In practice clusters are unlikely to be of equal size, and in that case the dependence of $\var(\widehat\theta)$ on the precise allocation of clusters to sequences will be complex. \citet{Kristunas} investigated corrections to the design effect for SWD which allow for unequal cluster sizes; \citet{girlinguneq} considered the case where clusters of sizes $n_1,\ldots,n_\alpha$ are allocated to each sequence; \citet{martinjt} investigated the effect on various aspects of the analysis of random allocation of clusters of arbitrary sizes to sequences.\\
The aim of the present paper is to identify the features of trial which determine the efficiency of an allocation and hence to identify efficient allocations.  The initial development will assume a cross-sectional design analyzed using the model of Hussey and Hughes, with the modifications necessary for the closed-cohort design being given subsequently. Specifically, it is assumed that $C$ clusters will be allocated to the $S$ sequences and that the number of individuals in a cluster-period cell of cluster $i$ is $N_i$.  Some period to period variation in the number recruited in cluster $i$ is inevitable but this is unlikely to be known at the planning stage and is ignored in the following.  It is helpful if the $N_i$ are known when the study is planned, but progress is possible if only estimates of their mean and dispersion are available.  It is also assumed that the number of periods, $T$, has been chosen and that a working estimate of $\rho$ is available.\\
When the numbers of clusters and sequences is modest an optimal design can be found by considering all possible allocations of clusters to sequences.  However, for larger $C$ and $S$ this rapidly becomes computationally infeasible.  The approach pursued in the present paper emphasises analytical over numerical methods, even when this requires some approximation, as the general form of such a design can provide greater insight than an isolated numerical solution. The issues encountered will be illustrated using two SWDs.  One is a four-period design being used in a trial of renal replacement therapy (RRT) in which the clusters are six intensive care units: the values of $N_i$ for this study are 6, 6, 6, 4, 4 and 2, with $\rho=0.1$.  The second is based on the Washington State Community-level Randomized Trial of Expedited Partner Therapy \citep{sexpart} for the treatment of partners of patients with sexually transmitted diseases, hereinafter the EPT trial.  This used a SWD with five periods and 22 clusters: the $N_i$ ranged from  91 to 1918, with a median of 272, and $\rho=0.0036$.  SWDs with these numbers of clusters are reasonably representative of the range of SWDs: a recent review \citep{Martinrev} found a median of 17 clusters per trial.\\
In Section~\ref{sect:calcs} a convenient expression is derived for $\var(\widehat\theta)$; Section~\ref{sect:optdes} derives optimal and highly efficient designs; Section~\ref{sect:numres} provides some numerical illustrations; Section~\ref{sect:practs} addresses the practical implications of the results; Section~\ref{sect:cohort} describes the extension to closed cohort designs and the resullts are discussed in Section~\ref{sect:disc}.\\
\section{Expression for the variance of $\widehat\theta$}\label{sect:calcs}
\subsection{Variance of $\widehat\theta$ in terms sequence allocations}\label{subsect:Ni}
The model of \citet{HusseyHughes} assumes that the outcome on individual $k=1,\ldots,N_i$ in period $j=1,\ldots,T$ of cluster $i=1,\ldots,C$, $Y_{ijk}$, is
\begin{equation}\label{eqn:model}
Y_{ijk}=\beta_j + X_{ij}\theta + \alpha_i +\epsilon_{ijk},
\end{equation}
where $\beta_j$ is the effect of period $j$ and $X_{ij}\in\{0,1\}$ is a treatment indicator which is 1 only if treatment B is administered to cluster $i$ during period $j$.  The cluster effects are modelled by the independent random variables $\alpha_i$ which have mean 0 and variance $\tau^2$, and the residual terms $\epsilon_{ijk}$ are independent of each other, and of the $\alpha_i$, and have mean 0 and variance $\iv$: consequently the ICC, $\rho=\tau^2/(\tau^2+\iv)$.  This model induces a model for the cluster-period means which can be written $\mathbf{Y}=Z\bm{\phi}+\mathbf{e}$, where $\mathbf{Y}^T=(Y_1,\ldots,Y_C)^T$ and $Y_i^T=(\overline{Y}_{i1+},\ldots,\overline{Y}_{iT+})^T$, where $\overline{Y}_{ij+}$ is the mean of the $Y_{ijk},k=1,\ldots,N_i$.  Here $Z$ is the $(CT)\times (T+1)$ design matrix for the induced model and $\bm{\phi}=(\beta_1,\ldots,\beta_T,\theta)^T$,  The dispersion matrix of  $\mathbf{e}$ is the $CT \times CT$ block diagonal matrix $V=\diag(V_1,V_2,\ldots,V_C)$, with $V_i=\var(Y_i)=\tau^2J_T+N_i^{-1}\iv I_T$, where $I_T$ is the $T \times T$ identity matrix and $J_T=1_T1_T^T$, with $1_T$ being the $T \times 1 $ vector of ones. Consequently $\var(\widehat\theta)$ is the bottom right hand element of $(Z^TV^{-1}Z)^{-1}$.\\
The search for a succinct and general expression for efficient designs is facilitated by expressing $\var(\widehat\theta)$ in terms of the proportions of subjects and other quantities allocated to the $S$ sequences of the design.  To this end define $p_i=N_i/N$, where $N=\sum N_i$, and $p_\ell=\sum_{i\in\ell} p_i$, where $\sum_{i\in\ell}$ denotes summation over all clusters allocated to sequence $\ell$, so that $p_\ell$ is the proportion of individuals allocated to sequence $\ell$. The vector of $p_\ell,\;\ell=1,\ldots,S$ is denoted by $P$. The vector $Q$ of the $q_\ell$ is analogously derived from $q_i$, which is defined by the equivalent forms 
\[
q_i=\frac{N_i^2}{N(\lambda+N_iT)}=p_i\frac{N_i}{\lambda+N_iT}=\frac{p_i^2}{\lambda'+p_iT},
\]
here $\lambda=(1-\rho)/\rho$ is a convenient transformation of the ICC and $\lambda'=\lambda/N$.  The quantity $W=\sum_{i=1}^C q_i$ plays an important role in the following and it should be noted that $0 \le W \le T^{-1}$.  With these definitions, Appendix A shows that 
\begin{equation}\label{eqn:objfun1}
N^{-1}\sigma_e^2\var(\widehat\theta)^{-1}=P^T\Xi P-Q^T\widetilde\Lambda P-\left (\frac{TQ^T\Delta Q+WP^T\Delta P-2Q^T\Delta P}{1-WT}  \right )
\end{equation}
where $\Xi,\widetilde\Lambda,\Delta$ are $S\times S$ matrices defined as follows.  Write $z$ for the vector of the integers 1 to $S$ centred about their mean, i.e. $z_\ell=\ell-\tfrac{1}{2}(S+1)$ and $y$ for the vector comprising the squares of the elements of $z$.  Then $\Delta=zz^T$, $\widetilde\Lambda=y1_S^T$, and $\Xi_{\ell,\ell'}=\tfrac{1}{2}\mid\ell-\ell'\mid$.\\
For both the EPT and RRT trials, and for a variety of simulated examples, the $q_i$ and $Wp_i$ exhibit almost a straight line relationship ($r>0.999$ for both EPT and RRT).   This relationship can be exploited to provide a very accurate and useful approximation to \eqref{eqn:objfun1}.\\  
\subsection{Regression-based approximation to $\var(\widehat\theta)$}
If we approximate the $q_i$ using the regression $q_i=\alpha+\beta Wp_i+r_i$, then the residuals $r_i$ will be small.  As $\sum q_i=W=\sum Wp_i$ and $\sum r_i=0$, $\alpha=C^{-1}W(1-\beta)$.  If the vector $R$ is formed from the $r_i$ as $P$ was formed from the $p_i$, the equation
\begin{equation}\label{eqn:regvec}
Q=W(1-\beta)K+\beta WP+R
\end{equation}
is obtained, where the $\ell$th element, $k_\ell$, of the vector $K$ is the proportion of \emph{clusters} allocated to sequence $\ell$.  If \eqref{eqn:regvec} is used to eliminate $Q$ from \eqref{eqn:objfun1}, and terms involving $R$ are neglected, then the following approximation is obtained:
\begin{equation}\label{eqn:mainapp}
N^{-1}\sigma_e^2\var(\widehat\theta)^{-1}\approx \mathscr{V}=\mathscr{V}(P,K)=P^TAP +h_1 b z^TP-h_2b^2-W(1-\beta)a 
\end{equation}
where $A=\Xi-\beta W \Lambda+\gamma W\Delta$, $\Lambda=\tfrac{1}{2}(\widetilde \Lambda+\widetilde\Lambda^T)$, and the constants $\gamma$, $h_1$ and $h_2$ are given by
\begin{align*}
h_1&=2W(1-\beta)\left ( \frac{1-\beta WT}{1-WT}\right )\qquad h_2=\frac{(1-\beta)^2 W^2T}{1-WT}\\
\gamma&=\frac{2\beta-1-\beta^2WT}{1-WT}.
\end{align*}
Details are in Appendix A, where it is also shown that $\beta>1$. 
The dependence of \eqref{eqn:mainapp} on the disposition of the clusters, $K$, is solely through the two linear forms $b=K^Tz$ and $a=K^Ty$, i.e.
\begin{align}
b=&\tfrac{1}{2}k_1(1-S)+\tfrac{1}{2}k_2(3-S)+\ldots+\tfrac{1}{2}k_S(S-1)\\
a=&\tfrac{1}{4}k_1(S-1)^2+\tfrac{1}{4}k_2(S-3)^2+\ldots+\tfrac{1}{4}k_S(S-1)^2\label{eqn:aks}.
\end{align}
The approximation \eqref{eqn:mainapp} is very accurate.  For the RRT trial there are 177 different allocations where $\theta$ is estimable.  For these allocations, with $\lambda=9$,  \eqref{eqn:mainapp} and \eqref{eqn:objfun1} differ by less than 1\% for 175 cases, with a discrepancy of 1.5\% only for two extreme allocations where five of the six clusters were allocated to a single sequence. For over 90\% of allocations the discrepancy was less then 0.5\%.  When $\lambda=19$ all but 4 are within 1\% and none differed by more than 2\%.   In only 14 of  $10^5$ unrestricted random allocation of the 22 clusters in the EPT trial did the exact and approximate values differ by more than 0.4\% and none differed by more than 0.6\% ($\lambda=276$): if $\lambda$ changed to 20, the values differed by less than 0.01\%.  When $\lambda=276$ and only allocations which allocate nearly equal numbers of clusters to each sequence (i.e. 5 clusters to two sequences and 6 to the other two sequences) are allowed, the exact and approximate forms differ by less than 0.35\%  and generally by less than 0.2\%.\\
An important symmetry property of SWDs should be noted. Two SWDs are said to be mirror images if the cluster allocated to sequence $\ell$ in one is allocated to $S+1-\ell$ in the other: the mirror image of $(P,K)$ is $(RP,RK)$, where $R$ is the matrix which reverses the order of a vector. \citet{kaszaforbes} showed that for models with centrosymmetric error structures, such as in \eqref{eqn:model}, $\var(\widehat\theta)$ is the same for mirror image designs.  This is preserved in \eqref{eqn:mainapp} because $\mathscr{V}(P,K)=\mathscr{V}(RP,RK)$.
\section{Optimal and Efficient designs}\label{sect:optdes}
\subsection{Cluster-balanced designs}
In many studies using SWDs the allocation of clusters to sequences is constrained so that equal numbers of clusters are allocated to each sequence.  There may be strong practical reasons for this.  For example, if the team training clusters in the new intervention visits different sequences in turn, then it may be impossible for them to accommodate widely varying numbers of clusters in their schedule.  Another reason is that balanced allocation of clusters preserves a degree of equity in when the cluster will receive the new intervention.  Among papers concerned with unequal clusters, both \citet{girlinguneq} and \citet{martinjt} assume that equal numbers of clusters are allocated to each sequence.\\
This assumption essentially fixes $K$, so that $a$ and $b$ in \eqref{eqn:mainapp} are fixed and maximization of $\mathscr{V}$ depends on the first two terms. If equal numbers of clusters are allocated to each sequence then $b=0$ and only the first term is pertinent.  The EPT trial allocated 22 clusters to 4 sequences and suitable symmetric allocation of the two `extra' clusters would also give $b=0$. However, had there been 23 clusters then it would not have been possible to arrange for $b$ to vanish and to accommodate this case the linear term in $P$ is retained for the moment.\\
The value of $P$ maximizing \eqref{eqn:mainapp} can be found explicitly as
\begin{equation}\label{eqn:fullopt}
P_\mathrm{opt}=W\beta 1_S + \tfrac{1}{2}(1-W\beta S)e-\frac{h_1b}{2[1-\gamma W(S-1)]}f,
\end{equation}
where $e,f$ are $S$-dimensional vectors of zeros, save for $e_1=e_S=1$ and $f_1=-f_S=1$: details are in Appendix B.\\
The optimal design allocates a proportion $W\beta$ of all the individuals in the trial to each of sequences $2,\ldots,S-1$.  If the allocation of clusters is such that $b=0$ then the remaining individuals are allocated equally to sequences 1 and $S$.  For non-zero $b$ the final term of \eqref{eqn:fullopt} creates an imbalance between the proportions allocated to the two outer sequences. \\
The optimum value of $\mathscr{V}$ can be written as $\mathscr{V}_\mathrm{opt}=(1^T_SA^{-1}1_S)^{-1}-h_3b^2-W(1-\beta)a$, where the form of $h_3$ is in Appendix B, where it is shown to be positive.  It follows that the maximum of $\mathscr{V}$ is larger if the clusters are arranged so that $b=0$, as would arise from equal or symmetrical allocation.\\
\subsection{Designs with $P=K$}
Although it will often be possible to obtain allocations with quite different $P$ and $K$, there are some inescapable links between these vectors.  In particular $p_\ell=0$ if, and only if, $k_\ell=0$.  One way this restriction can be accommodated is to consider allocations in which the aim is to allocate clusters and individuals to sequences in equal proportions, i.e. take $P=K$.  This case covers designs where all the clusters have the same size, $n$ and also the case envisaged in \citet{girlinguneq}.\\
When $P=K$ the expression for $\mathscr{V}$ simplifies to 
\begin{equation}\label{eqn:casePisK}
\mathscr{V}(P,P)=P^T(\Xi-W\Lambda+W\Delta)P,
\end{equation} 
and the allocation maximizing $\mathscr{V}(P,P)$ is $W1_S+\tfrac{1}{2}(1-WS)e$.  If all clusters are of size $n$ then $W=n/(\lambda+nT)$, which is the result of \citet{forbesopt}.\\
When equal numbers of clusters and individuals are allocated to every sequence, i.e. $P=K=\tfrac{1}{S}1_S$, the value of $\mathscr{V}$ is $S^{-2}(1^T_S\Xi 1_S-W1^T_S\Lambda 1_S)$, because $\Delta 1_S=0$.  As $1^T_S\Lambda 1_S=Sy^T1_S>0$, $\var(\widehat\theta)$ increases with $W$.  In \citet{girlinguneq} clusters of sizes $n_1,\ldots,n_\alpha$ were allocated to each sequence, so $NW=S\sum_{h=1}^\alpha n^2_h(\lambda+n_hT)^{-1}$.  If it is assumed that $\sum n_h$ is held fixed, then standard methods show that $W$, and hence $\var(\widehat\theta)$, are minimized when the $n_h$ are all equal, as observed by \citet{girlinguneq}.
\subsection{Unrestricted allocations}
If there is no restriction on how many clusters might be allocated to each sequence, then an optimal design results from maximizing \eqref{eqn:mainapp} over both $P$ and $K$.  However, not every allocation of individuals, $P$, will be compatible with every allocation of clusters $K$, and this will limit the amount of progress that can be made in general.  However, the form of \eqref{eqn:mainapp} permits sensible choices to be made.\\
Evaluation of $1_S^TA^{-1}1_S$ shows that for fixed $b$ and $a$, $\mathscr{V}_\mathrm{opt}$ is 
\begin{equation}\label{eqn:maxV}
\frac{1}{12}(S-1)(3-3(S-1)W\beta+S(S-2)W^2\beta^2)-h_3b^2-W(1-\beta)a.
\end{equation}
Focussing on allocations with $b=0$ and recalling that $W(1-\beta)<0$, it can be seen from \eqref{eqn:aks} that this term will cause $\mathscr{V}$ to increase if more clusters are allocated to the outer sequences.  The increase in this term depends only on the number of clusters moved to outer sequences, not on their sizes.  However, there are limits to how far this process can be continued; while the term $-W(1-\beta)a$ would in itself be maximized by allocating all the clusters to sequences 1 and $S$, this would not be compatible with the $P_\mathrm{opt}$ in \eqref{eqn:fullopt} unless $W\beta=0$.  \\
If there is some flexibility in how $P_\mathrm{opt}$ is obtained, then exercising this in such a way that more clusters are allocated to the outer sequences would be beneficial.  If there are some small clusters then it may be possible to allocate these in this way without making noticeable changes to $P_\mathrm{opt}$.  More fundamentally, there may be circumstances where a sub-optimal $P$ and a larger value of $a$ combine to produce a design not of the form \eqref{eqn:fullopt} which has a larger $\mathscr{V}$.  However, this issue could only really be satisfactorily resolved by knowledge of which vectors $P$ and $K$ can be obtained for a given set of clusters $N_1,\ldots,N_C$, so a general resolution is likely to be hard to achieve. In the next section some numerical results illustrate these points and show that in most cases ensuring that $P$ has the correct form is more important than the detailed disposition of the clusters.\\
\section{Some numerical results and simulations}\label{sect:numres}
In this section the implications of the results derived in Section \ref{sect:optdes} will be illustrated, largely using the RRT and EPT trials.  However, an artificial example will provide a clearer understanding of the roles of the allocations of individuals and of clusters.\\
Consider a SWD with $T=5$ and eight clusters, four of size 20 and four of size 10 and suppose that $\lambda=50$.  This arrangement gives $W=\tfrac{11}{90}$ and $\beta=\tfrac{15}{11}$ and hence $W\beta=\tfrac{1}{6}$. Let $D1$ and $D2$ be the two allocations shown in Table \ref{tab2}.
\renewcommand{\arraystretch}{1.2}
\begin{table}
\begin{center}
\caption{A SWD for $S=4,T=5$, with two allocations, $D1$ and $D2$, of eight clusters, four of size 10 and four of size 20.}
\label{tab2}
\begin{tabular}{cc|cccccccc}
Allocation $D1$& Allocation $D2$&\multicolumn{5}{c}{Periods}&$P$&$K\;(D1)$&$K\;(D2)$\\
\hline
20 20&20 10 10& A & A& A& A& B&$\tfrac{1}{3}$&$\tfrac{1}{4}$&$\tfrac{3}{8}$ \\
10 10&20 & A & A& A& B& B&$\tfrac{1}{6}$&$\tfrac{1}{4}$&$\tfrac{1}{8}$ \\
10 10&20 & A & A& B& B& B&$\tfrac{1}{6}$&$\tfrac{1}{4}$&$\tfrac{1}{8}$ \\
20 20&20 10 10& A & B& B& B& B&$\tfrac{1}{3}$&$\tfrac{1}{4}$&$\tfrac{3}{8}$ \\
$\mathscr{V}_{D1}=0.486$&$\mathscr{V}_{D2}=0.508$&&&&&&&&
\end{tabular}
\end{center}
\end{table}
With the allocations of clusters fixed at either $K=(\tfrac{1}{4},\tfrac{1}{4},\tfrac{1}{4}
,\tfrac{1}{4})^T$ ($D1$) or at $K=(\tfrac{3}{8},\tfrac{1}{8},\tfrac{1}{8}
,\tfrac{3}{8})^T$, ($D2$), then $b=0$ and the optimal $P$ would be $(\tfrac{1}{3},\tfrac{1}{6},\tfrac{1}{6}
,\tfrac{1}{3})^T$.  While $D1$ and $D2$ share this value of $P$,  $D2$ has a larger $\mathscr{V}$ because more clusters are allocated to sequences 1 \& 4, which gives a larger value of $a$ (1.25 for $D1$ and 1.75 for $D2$), so the term $-W(1-\beta)a$ in \eqref{eqn:mainapp} is larger. The difference between $\mathscr{V}_{D1}$ and $\mathscr{V}_{D2}$ varies with $\lambda$, with the difference being greatest for $\lambda$ in the region 50 to 70, reducing to zero when $\lambda$ is 0 or infinite: see Appendix C for more details.  This pattern arises because when $\lambda\rightarrow\infty$ the members of a cluster are independent, so the way the individuals are divided into clusters is immaterial.  When $\lambda$ is small the term $\alpha_i$ has a relatively high variance, so the estimator for $\theta$ will essentially be based on within-cluster contrasts, which largely eliminate $\alpha_i$ \citep[see][]{matthewsforbes}, which is why the clustering of the individuals is also unimportant when $\lambda$ is small. \\ 
\subsection{Cluster-balanced designs}\label{sect:clusbal}
\subsubsection{The EPT trial}\label{sect:clusbalEPT}
\begin{figure}
\centering
\includegraphics[width=0.9\linewidth]{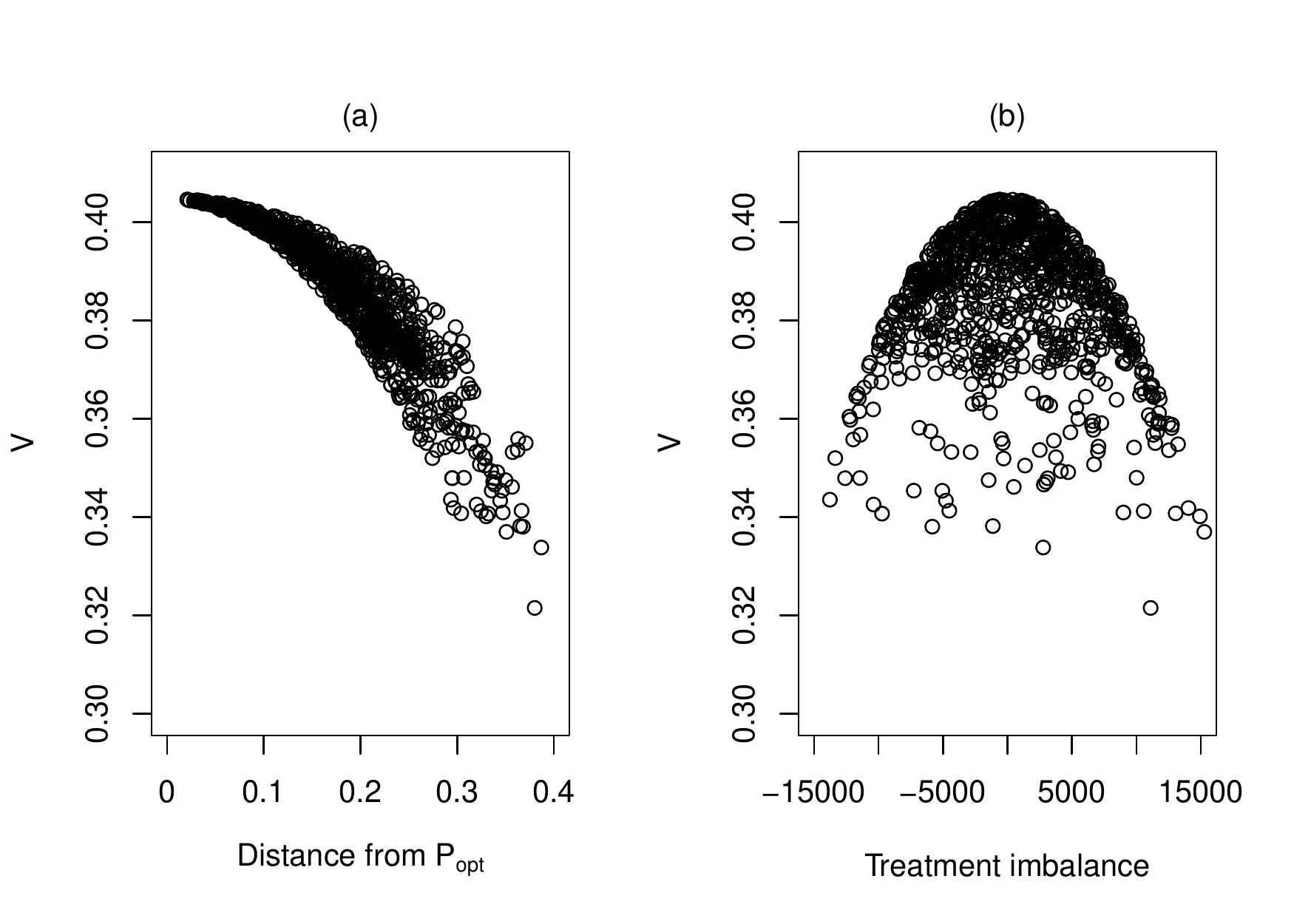}
\caption{(a) $\mathscr{V}$ vs. distance of the allocated $P$ from the optimum, (b) $\mathscr{V}$ plotted against imbalance in numbers in A and B (negative if more on A): for clarity only a random selection of 1000 of the $10^6$ cluster-balanced allocations for the EPT trial with $\lambda=276$ are shown.}\label{fig:newfig1}
\end{figure}

If allocation of the 22 clusters in the EPT trial is such that the number of clusters on each sequence is to differ as little as possible,
 then five clusters must be allocated to each sequence, with two extra clusters .  Consideration of \eqref{eqn:mainapp} suggests allocating the extra clusters symmetrically, i.e. to 1 \& 4 or to 2 \& 3, so as to maintain $b=0$.  Also, all else being equal, allocation to 1 \& 4 would make $a$ larger than allocation to 2 \& 3, so six clusters were allocated to sequences 1 and 4, with five each to sequences 2 and 3.\\
For the EPT trial, with $\lambda=276$ ($\rho=0.0036$), $W=0.1816$, $\beta=1.0900$ and $W\beta=0.1980$.  For this allocation of clusters the maximum $\mathscr{V}$ according to \eqref{eqn:maxV} is 0.4048342.  The largest $\mathscr{V}$ observed in $10^6$ random allocations with this pattern of clusters was 0.4048339: calculations were done in \R, \citep{Rcite}.  The optimal allocation of individuals to clusters according to \eqref{eqn:fullopt} is $P_\mathrm{opt}=(\tfrac{1}{2}-W\beta,W\beta,W\beta,\tfrac{1}{2}-W\beta)^T=(0.302,0.198,0.198,0.302)^T$.  The allocation which maximized $\mathscr{V}$ over all $10^6$ allocations deviated from $P_\mathrm{opt}$ only by rounding error.  \\
The effect of the allocated $P$ deviating from $P_\mathrm{opt}$ is shown in Figure~\ref{fig:newfig1}, where the distance $\sqrt{(P-P_\mathrm{opt})^T(P-P_\mathrm{opt})}$ is plotted against $\mathscr{V}$ (for clarity only a random selection of 1000 allocations is shown).  It can be seen that $\mathscr{V}$ decreases as the distance from the optimum increases: for larger deviations $\mathscr{V}$ can be reduced by 20\%, although the reduction is more modest for smaller deviations. \\
\citet{martinjt} investigated the effect on power of the imbalance in the number of individuals on treatments A and B.  In the present notation this is $2NP^Tz$ and there is no imbalances with the optimal design because $P_\mathrm{opt}^Tz=0$. The right hand panel of Figure~\ref{fig:newfig1} confirms the findings in \citet{martinjt}, that designs which minimize the imbalance are to be preferred.\\
The results for $\lambda=20$ are not shown because they are very similar to those for $\lambda=276$.\\
\subsubsection{The RRT Trial}\label{sect:restrrt}
For the RRT trial with $\lambda=9$ ($\rho=0.1$), $W=0.1710$, $\beta=1.2644$, $W\beta=0.2162$, and $P_\mathrm{opt}=(0.39,0.22,0.39)^T$.  The corresponding values when $\lambda=19$ ($\rho=0.05$) are $W=0.1276$, $\beta=1.3774$, $W\beta=0.1758$, giving $P_\mathrm{opt}=(0.41,0.18,0.41)^T$. With two clusters allocated to each sequence the maximum possible value for $\mathscr{V}$ is, from \eqref{eqn:maxV}, 0.3373 ($\lambda=9$) and 0.3717 ($\lambda=19$).  \\
As there are only six clusters and three sequences an allocation can be written succinctly as, e.g. (6,2;6,6;4,4), meaning that clusters of size 6 and 2 were allocated to sequence 1, two clusters of size 6 were allocated to sequence 2 and two clusters of size 4 to sequence 3.  There are 15 cluster-balanced allocations and the maximum $\mathscr{V}$ over these allocations is 0.3360 ($\lambda=9$) and 0.3695 ($\lambda=19$), both over 99\% of the values from \eqref{eqn:maxV}.  The difference is due to the $P$ for the observed maxima, namely $P=(0.357,0.214,0.429)^T$ for both values of $\lambda$, not coinciding with $P_\mathrm{opt}$.  The designs with maximum $\mathscr{V}$ correspond to (6,4;4,2;6,6) or its mirror image (6,6;4,2;6,4).  With such a small number of clusters it is not possible to find an allocation giving $P_\mathrm{opt}$, although the observed maximum corresponds to an allocation which minimizes the distant of the observed $P$ from $P_\mathrm{opt}$. \\
\subsection{Unrestricted allocations}
\subsubsection{Unrestricted allocation of clusters to sequences in the EPT trial}
\begin{figure}
\centering
\includegraphics[width=0.9\linewidth]{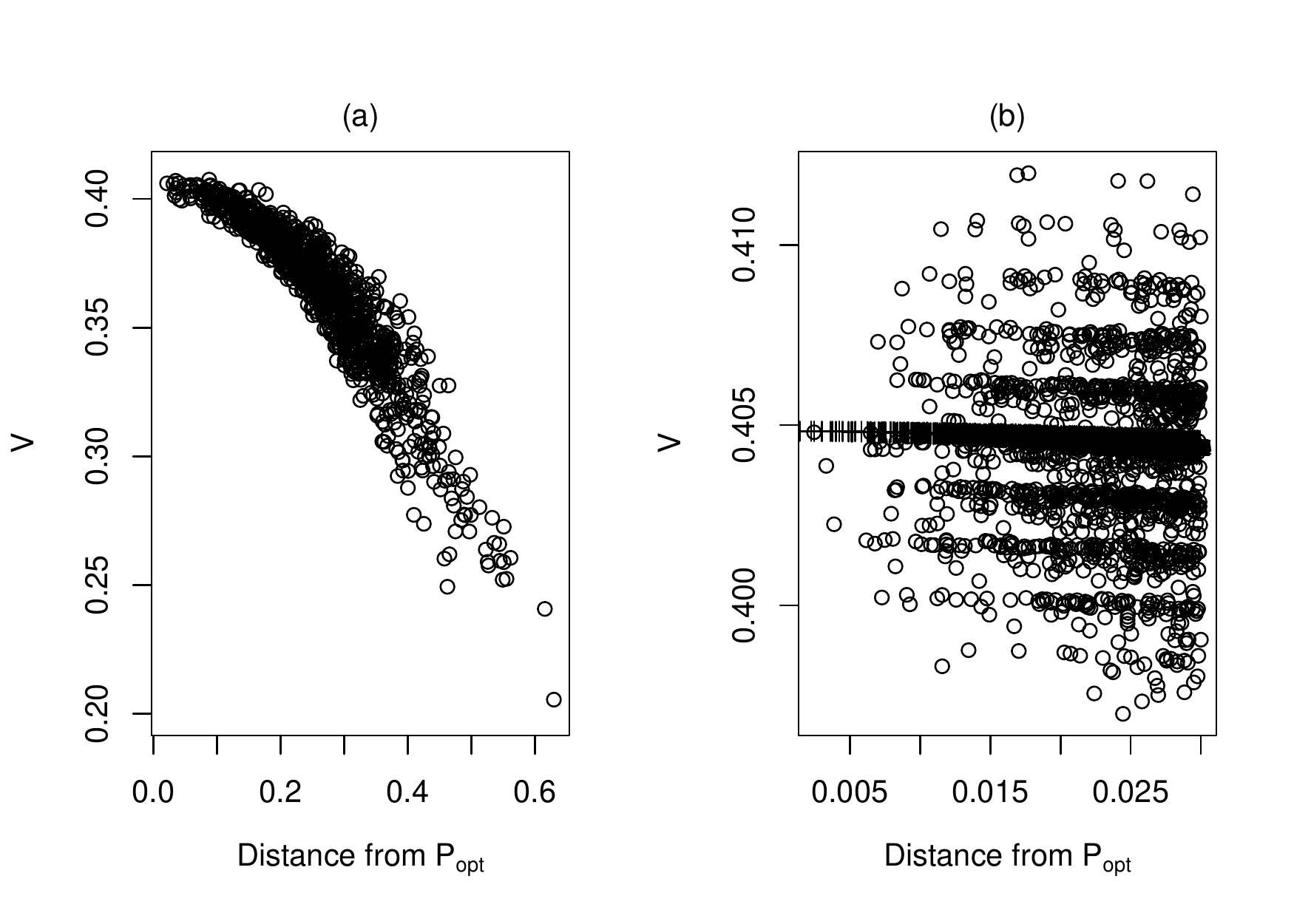}
\caption{(a) $\mathscr{V}$ vs. distance of the allocated $P$ from the optimum, for a random selection of 1000 of the $10^6$ unrestricted allocations for the EPT trial with $\lambda=276$: (b) same as (a) but showing all allocations with distance less than 0.03 ($\ocircle$); the corresponding points for the cluster-balanced allocation are shown for comparison (+)}\label{fig:newfig2}
\end{figure}

A million random allocations were made of the 22 clusters to the 4 sequences of EPT trial, without any restriction on the number of clusters allocated to each sequence.  The contributions of the various terms in \eqref{eqn:mainapp} to $\mathscr{V}$ are shown in Table \ref{tab4}.  It is clear that with the values of $W$ and $W\beta$  for the EPT trial the second and third terms, which involve $b$, make little contribution to $\mathscr{V}$.  While the term involving $a$ is not wholly negligible, by far the main contribution to $\mathscr{V}$ is from the first term, which depends only on the proportions of individuals allocated to each sequence.\\
\begin{table}
\begin{center}
\caption{Mean values of the absolute values of the terms in \eqref{eqn:mainapp} for $10^6$ unrestricted random allocations in the EPT trial. Values are shown for $\lambda=276,20$ and for the means of all allocations and the 1000 allocation with largest $\mathscr{V}$.}
\label{tab4}
\begin{tabular}{ll|ccccc}
&&$\mathscr{V}$&$P^TAP$ & $\mid h_1 b z^TP\mid$&$\mid h_2b^2\mid$&$\mid W(1-\beta)a\mid$\\
$\lambda=276$&All allocations&0.3637& 0.3442 &0.0002& 0.0008& 0.0204\\
&Largest 1000& 0.4085& 0.3814& 0.0000& 0.0003& 0.0274\\
$\lambda=20$&All allocations&0.3437& 0.3418& 0.0000 &0.0001& 0.0020\\
&Largest 1000&0.3820& 0.3797& 0.0000& 0.0000& 0.0023
\end{tabular}
\end{center}
\end{table}
The optimum $P$ for fixed $a$ and $b=0$ is the same as the optimum for the cluster-balanced designs.  Using this value for $P_\mathrm{opt}$, the distance of each $P$ from $P_\mathrm{opt}$ can be found as before and this distance is plotted against $\mathscr{V}$ in Figure~\ref{fig:newfig2}.  As in Figure~\ref{fig:newfig1} the main determinant of $\mathscr{V}$ is the distance of the allocation of individuals from $P_\mathrm{opt}$.  The right hand panel in Figure~\ref{fig:newfig2} focusses on those allocations where the distance is less than 0.03 (from all $10^6$ allocations, i.e. with no selection).  Here it is seen that the optimal design is not the one which minimizes the distance from $P_\mathrm{opt}$, because near the optimum allocations putting some small clusters on the extreme sequences may be more beneficial than minimizing distance from $P_\mathrm{opt}$.  However, the important point is that the variation on the vertical axis in the right hand panel of Figure~\ref{fig:newfig2} is very small.  Figure~\ref{fig:newfig2} also shows the corresponding points for the cluster-balanced allocation, where optimality is determined solely by minimizing distance from $P_\mathrm{opt}$.  It should also be noted that in this trial the gain in $\mathscr{V}$ of allowing unrestricted allocation has been very small.\\
For $\lambda=276$ the maximum $\mathscr{V}$ obtained from the $10^6$ allocations is 0.4129, when $P=(0.280,0.232,0.165,0.322)^T$ and $K=(0.409,0.091,0.091,0.409)^T$.  A higher proportion of clusters than individuals has been allocated to sequences 1 and 4.  In the EPT trial the mean cluster size is 495.2.  With allocation $(P,K)$ the mean cluster size in sequence $\ell$ is $(Np_\ell)/(Ck_\ell)$, which is 339, 1262, 898, 390 for sequences 1 to 4 respectively.  So the allocation appears to have obtained good $P,K$ by allocating fewer, larger clusters to the inner two sequences.  The mean cluster sizes for the optimum cluster-balanced allocation are 548, 429,431, 548, so the discrepancy in mean cluster size is more exaggerated in the unrestricted case.  For $\lambda=20$ very similar results are found.

\subsubsection{Unrestricted allocation of clusters to sequences in the RRT trial}\label{sect:unrrrt}
The values of $W$ and $\beta$ for $\lambda=9$ and $19$ are as in Section \ref{sect:restrrt}.  The value of $\mathscr{V}$ was computed for all 177 possible allocations.  For $\lambda=9$ the allocation which maximizes $\mathscr{V}$ is (4,4,2;6;6,6), and has $P=(0.357,0.214,429)^T$ and $K=\tfrac{1}{6}(3,1,2)^T$: the mirror-image design (6,6;6;4,4,2) also maximizes $\mathscr{V}$.  \\
The designs with the four largest $\mathscr{V}$ are shown in Table \ref{tab5} (excluding mirror-image designs), for both $\lambda=9$ and $\lambda=19$.  The three allocations with the largest $\mathscr{V}$ for each $\lambda$ all have  $b=-\tfrac{1}{6}$ so from \eqref{eqn:fullopt} $P_\mathrm{opt}=(0.386,0.216,0.398)^T$ is the same for these designs.  All these allocations put only one cluster on sequence 2, which results in a value for $-W(1-\beta)a$ that is exceeded only by designs which do not use sequence 2 (cf. allocation 8).  The vector $P$ in allocations 1 and 2 are mirror images, whereas $K$ is unchanged in the two cases, so $h_1bz^TP$ has the same absolute value in the two cases but is positive in the former and negative in latter, which is why allocation 1 is ahead of allocation 2. For allocation 3, $P$ is further from $P_\mathrm{opt}$, giving a smaller first term in \eqref{eqn:mainapp} than in allocations 1 \& 2.\\
\begin{landscape}
\begin{table}
\begin{center}
\caption{Four allocations with largest $\mathscr{V}$ for the RRT trial, for $\lambda=9,19$. Distance denotes distance between $P$ and $P_\mathrm{opt}$. $P$ and $K$ are multiplied by $N$ and $C$ respectively for clarity. The individual terms from \eqref{eqn:mainapp} are in the final four columns.}
\label{tab5}
\begin{tabular}{llcccccccc}
&Allocation & $\mathscr{V}$ & Distance& $28P^T$& $6K^T$& $P^TAP$ & $ h_1 b z^TP$&$ -h_2b^2$&$-W(1-\beta)a$\\
\hline
\multicolumn{10}{c}{$\lambda=9$}\\
\hline
1&4,4,2;6;6,6 & 0.343& 0.042&(10,6,12)&(3,1,2)&0.306&0.0005&-0.0007&0.038\\
2&6,4,2;6;6,4 & 0.342& 0.059&(12,6,10)&(3,1,2)&0.306&-0.0005&-0.0007&0.038\\
3&6,4,2;4;6,6 & 0.341 & 0.090&(12,4,12)&(3,1,2)&0.304&0&-0.0007&0.038\\
4&6,4;4,2;6,6&0.336&0.051&(10,6,12)&(2,2,2)&0.306&0&0&0.030\\
\hline
\multicolumn{10}{c}{$\lambda=19$}\\
\hline
5&6,4,2;4;6,6 & 0.379& 0.042&(12,4,12)&(3,1,2)&0.339&0&-0.0005&0.040\\
6&4,4,2;6;6,6 & 0.378& 0.061&(10,6,12)&(3,1,2)&0.337&0.0007&-0.0005&0.040\\
7&6,4,2;6;6,4&0.376&0.078&(12,6,10)&(3,1,2)&0.337&-0.0007&-0.0005&0.040\\
8&6,6,2;;6,4,4&0.372&0.215&(14,0,14)&(3,0,3)&0.324&0&0&0.048\\
\hline
\end{tabular}
\end{center}
\end{table}
\end{landscape}
For $\lambda=19$ $P_\mathrm{opt}=(0.404,0.176,0.420)^T$, so the $P=\tfrac{1}{28}(12,4,12)^T$ in allocation 5 is now closer to $P_\mathrm{opt}$ than it was in allocation 3, and this individually symmetric allocation is optimal.  Allocation 8 is the only allocation in Table \ref{tab5} that is symmetric in both $P$ and $K$.  However, while putting all clusters on the outer sequences maximizes $-W(1-\beta)a$, it entails a value of $P$ that is too far from $P_\mathrm{opt}$ for the larger final term to compensate for the reduction in $P^TAP$.\\
In only two cases (allocations 4 \& 8) is $b=0$, and only allocation 4 is a cluster-balanced design.  For $\lambda=19$ the design with the seventh largest $\mathscr{V}$ is the best cluster-balanced design with $\mathscr{V}=0.370$.  However, while cluster-balanced designs may not be optimal in the RRT trial, the best designs of this type are more than 97\% efficient.
\section{Some practical considerations}\label{sect:practs}
The results presented in Figure~\ref{fig:newfig2} and Section~\ref{sect:unrrrt} suggest that cluster-balanced designs lose little in terms of efficiency.  Taken with the practical advantages that will often obtain, it is likely that many practitioners will prefer this form of allocation.  While it is straightforward to allocate clusters randomly so that the numbers on each sequence are as close to equal as possible, it is less easy to arrange matters so that the proportion of individuals, $P$, approximates the optimal form.  Moreover, even if an algorithm were available to generate the $P$ closest to $P_\mathrm{opt}$, many users would still wish to include some element of randomization in their choice of design.  The use of the foregoing results to avoid inefficient designs may be more attractive than focussing on a single optimum.\\
If the sizes of the clusters are known \emph{a priori}, then $W$ and $\beta$ can be evaluated.  A practitioner could generate a number of cluster-balanced random allocations and then use $W$ and $\beta$ and \eqref{eqn:maxV} to compare each $\mathscr{V}$ with its maximum possible value.  As an illustration, 1000 random allocations of the 22 clusters in the EPT trial were generated, with six clusters allocated to each of sequences 1 \& 4 and five to each of 2 \& 3. Using the values of $W$ and $\beta$ in Section~\ref{sect:clusbalEPT}, the ratio of $\mathscr{V}$ to $\mathscr{V}_\mathrm{opt}$ was calculated for each allocation.  One hundred of these allocations had efficiency less than 90\%, 392 less than 95\%, 770 less than 98\% and 910 less then 99\%.  Randomly choosing one of the 90 allocations with efficiency of at least 99\% would represent a sensible compromise between ensuring efficiency, while retaining randomness in the selection of the design.\\
If the values of the $N_i$ are not known when the trial is planned, then some progress in estimating $W$ and $W\beta$ can be made provided values are available for the mean, $M$, and coefficient of variation, $CV$, of the cluster sizes.  Approximations to $W$ and $W\beta$, found using the delta method, are
\begin{align}
\expect(W)&\approx \frac{M}{\lambda+MT}+\frac{\lambda^2M}{(\lambda+MT)^3}CV^2\nonumber\\
\expect(W\beta)&\approx \frac{1}{T}\left ( 1- \frac{\lambda^2}{(\lambda+MT)^2}\right).\label{eqn:wbetaapp}
\end{align}
Details are given in Appendix C.  Note that, to within the level of approximation, \eqref{eqn:wbetaapp} implies that $0\le W\beta \le T^{-1}$, so $P_\mathrm{opt}$ will always allocate at least 50\% more individuals to an outer sequence than to an inner one.\\
\section{Extension to closed-cohort SWDs}\label{sect:cohort}
In the closed-cohort SWD each individual is measured in every period, so $Y_{i1k},\ldots,Y_{iTk}$ are the successive measurements on individual $k$ in cluster $i$.  To accommodate the dependence between these observations additional, independent random effects $\zeta_{ik}$, independent of the other random effects and with zero mean and variance $\omega^2$, are included in \eqref{eqn:model}.  The vector $Y_i$ now has variance $V_i=(\tau^2+N_i^{-1}\omega^2)J_T+N_i^{-1}\iv I_T$.  The methods outlined in Sections~\ref{sect:calcs} and \ref{sect:optdes} can be applied to the closed-cohort design provided $W$ is replaced by $\widetilde W=\sum \widetilde q_i$, where
\[
\widetilde q_i=p_i\frac{N_i+\mu}{\lambda+T(N_i+\mu)},
\]
with $\mu=\omega^2/\tau^2$, and $\beta$ is replaced by $\widetilde\beta$, found from the regression of $\widetilde q_i$ on $\widetilde W p_i$.  Details are available in Appendix D. \\
If $N_i=n$ for all $i$, then $P=K$ and the optimum design allocates a proportion of clusters $\widetilde W=\sum_{i=1}^C\widetilde q_i=(n+\mu)/[\lambda+T(n+\mu)]$ to each of sequences 2 to $S-1$, with the remaining clusters equally allocated to the outer sequences: this is the result in \citet{Lietal}.
\section{Discussion}\label{sect:disc}
The analysis presented in this paper shows that efficient SWDs can be obtained provided that both the proportion of individuals, $P$, and the proportion of clusters, $K$, are appropriately allocated to the sequences of the design.  Cluster-balanced designs, in which clusters are allocated equally to sequences are likely to have practical appeal and can have good statistical properties. For these designs it is optimal to allocate the same proportion, $W\beta$, of individuals to sequences 2 to $S-1$, with the remainder to the outer sequences.  In most cases the allocation to the outer sequences will be done equally to 1 and $S$, with any imbalance being for cases where $b\ne 0$ and even then it will be small.  When $\rho=0$, i.e. all observations are independent, $W\beta=0$, so sequences 2 to $S-1$ are omitted.  The reason for this is that with independent observations a fully efficient design which eliminated period effects would be a randomized block design with periods as blocks. Using just sequences 1 and $S$ provides the closest approximation to such a design that can be obtained from the sequences of the SWD.  When $\rho=1$, the estimator of $\theta$ will be based solely on within-cluster contrasts.  In this case a proportion $T^{-1}$ of individuals is allocated to the inner sequences, with $3(2T)^{-1}$ to the outer: in this case an intuitive explanation for the form of the design is more elusive.\\
\citet{martinjt} explored cluster-balanced SWDs with unequal cluster sizes and noted that, for a fixed total number of individuals, SWDs with unequal cluster sizes could be more efficient than designs with equal cluster sizes - in contrast to traditional cluster-balanced designs where clusters of equal size are optimal.  A cluster-balanced SWD design with clusters of equal size will inevitably allocate equal proportions of individuals to the sequences, however with unequal cluster sizes some allocations may result in a design where the proportions of individuals on the sequences match the optimal allocation more closely than would be possible when cluster sizes do not vary.\\
When the proportion of clusters can vary between sequences, matters are less clear cut: \eqref{eqn:mainapp} indicates that more clusters should be allocated to the outer sequences, provided that in doing so the ability to obtain an allocation of individuals close to $P_\mathrm{opt}$ is not seriously compromised. While such designs may be better than cluster-balanced designs, the examples in the present paper suggest that this advantage is likely to be minor.  It is possible that the difference may be greater for larger $T$, when $a$ could be larger, but in such cases it should be remembered that the upper bound on $W$ and $W\beta$, $T^{-1}$, will be smaller.\\
A weakness of focussing on optimal designs is that they seek to optimise one quantity, here $\var(\widehat\theta)$, and this can lead to designs with a form, such as omitting sequences 2 to $S-1$, that would be unappealing.  It will often be sensible to use the results on optimality to avoid inefficient designs, as outlined in Section~\ref{sect:practs}, where the benefits of randomization are also retained.\\
It is unclear to what extent the detailed mathematics of the current approach would carry over to other error structures.  It is unlikely that the method would apply to the case with autocorrelated errors \citep{kasza1}, as it relies on the availability of an explicit inverse for $V_i$.  However, an advantage of the current approach is the succinctness of the description of an efficient design by the sets of proportions $P$ and $K$.  A numerical method which attempted to characterise efficient designs similarly, but which could accommodate a wider range of models, would have merit.

\section*{Acknowledgements}
I am grateful to Professor Jim Hughes for providing data and information from the EPT trial.  

\bibliographystyle{biom}
\bibliography{swdes}

\newpage
\renewcommand{\thefigure}{A\arabic{figure}}
\renewcommand{\thetable}{A\arabic{table}}
\renewcommand{\theequation}{A-\arabic{equation}}

\setcounter{figure}{0}
\setcounter{table}{0}
\setcounter{equation}{0}

\section*{Appendices}

Note that equations defined in the following appendices are of the form (A-$n$) - equation numbers without the A- prefix refer to equation numbers in the main part of the paper. 

\section*{Appendix A: {\large evaluating $\var(\hat\theta)$}}

\subsection*{{\large Evaluation of bottom right-hand element of $(Z^TV^{-1}Z)^{-1}$}}

In this section the matrix manipulations needed to evaluate $\var(\hat\theta)$ are presented.

The matrix $V_i=\tau^2 J_T+\iv N_i^{-1} I_T$, so 
\[
V_i^{-1}=\frac{N_i}{\iv}(I_T-w_iJ_T), 
\]
where $w_i=N_i\tau^2/(\iv+N_i\tau^2T)=N_i/(\lambda+N_iT)$.  Therefore $\sum V_i^{-1}=\sigma_e^{-2}N(I_T-WJ_T)$ and from this it follows that 
\[
(\sum V_i^{-1})^{-1}=\frac{\iv}{N}(I_T+\frac{W}{1-WT}J_T).
\]
The evaluation of $\var(\hat\theta)$ proceeds by noting that  
\[
Z=
\begin{pmatrix}
I_T & D_1\\
I_T & D_2\\
\vdots & \vdots \\
I_T & D_C\\
\end{pmatrix},
\]
where $D_i$ is the $T \times 1$ vector of 0s and 1s indicating the treatment allocation for cluster $i$.  It follows that 
\[
(Z^TV^{-1}Z)^{-1}=\begin{pmatrix}
\sum V_i^{-1} & \sum V_i^{-1}D_i\\
\sum D_i^TV_i^{-1} & \sum D_i^TV_i^{-1}D_i
\end{pmatrix}^{-1},
\]
and therefore using standard expression for inverting partitioned matrices
\[
\var(\hat\theta)^{-1}=\sum D_i^TV_i^{-1}D_i - (\sum D_i^TV_i^{-1})(\sum V_i^{-1})^{-1}(\sum V_i^{-1}D_i).
\]
Each term in this expression is now evaluated.

\subsubsection*{Evaluating $(\sum D_i^TV_i^{-1})(\sum V_i^{-1})^{-1}(\sum V_i^{-1}D_i)$}

If we denote by $r_i$ the number of times treatment B appears in the sequence to which cluster $i$ is allocated, then $r_i=D_i^TD_i=1^T_TD_i$.  Consequently $V_i^{-1}D_i=\sigma_e^{-2}N_i(D_i-r_iw_i1_T)$ and therefore
\[
\sum_{i=1}^C V_i^{-1}D_i=\sigma_e^{-2}\left \{ \sum_{i=1}^C N_iD_i -\left(\sum_{i=1}^C N_ir_iw_i\right)1_T \right\}.
\]
It follows that $(\sum V_i^{-1})^{-1}(\sum V_i^{-1}D_i)$ is
\[
\frac{1}{N} \left (\sum_{i=1}^C N_iD_i - M1_T \right )
\]
where
\[
M=\frac{1}{1-WT}\sum_{i=1}^CN_ir_iw_i-\frac{W}{1-WT}\sum_{i=1}^CN_ir_i=\frac{F-WE}{1-WT},
\]
and we have written $E=\sum_{i=1}^C N_ir_i$ and $F=\sum_{i=1}^C N_iw_ir_i$.  It follows that
\[
(\sum D_i^TV_i^{-1})(\sum V_i^{-1})^{-1}(\sum V_i^{-1}D_i)=N^{-1}\sigma_e^{-2}\left( G+\frac{TF^2+WE^2-2EF}{1-WT}\right)
\]
where $G=(\sum N_iD_i^T)(\sum N_iD_i)$.

\subsubsection*{Evaluating $\sum D_i^TV_i^{-1}D_i$}

\begin{align}
\sum_{i=1}^C D_i^TV_i^{-1}D_i&=\sigma_e^{-2}\sum_{i=1}^C N_iD_i^T(I_T-w_iJ_T)D_i\\&
=\sigma_e^{-2}\left(\sum_{i=1}^C N_ir_i-\sum_{i=1}^C N_iw_ir_i^2\right)\\
&=\sigma_e^{-2}(E-H),
\end{align}
with $H=\sum_{i=1}^C N_iw_ir_i^2$.\\

Putting the above together we obtain
\[
\sigma_e^2\var(\hat\theta)^{-1}=E-H-\frac{G}{N}-\frac{TF^2+WE^2-2EF}{N(1-WT)}.
\]
\subsection*{$\var(\hat\theta)$ in terms of sequence allocations} 

The clusters are to be allocated to the $S$ sequences in Table \ref{tab1} of the main paper.  If cluster $i$ is allocated to sequence $\ell$ then for all such clusters $r_i=r_\ell$, where $r_\ell$ denotes the number of occurrences of B in sequence $\ell$, which means that the quantities $E,F,G,H$ above can be re-written in terms of sums over sequences instead of sums over clusters. Note that with the sequences numbered as in Table \ref{tab1} $r_\ell=\ell$ and $r_{\ell\ell'}=\min\{\ell,\ell'\}$, $\ell,\ell'=1,\ldots,S$,\\

It is convenient to have an expression for $\var(\hat\theta)$ that is in terms of summations over sequences rather than cluster and this can be achieved as follows.
Define $p_i,q_i$ and $p_\ell,q_\ell$ as in the main paper, $i=1,\ldots,C$ and $\ell=1,\ldots,S$.  These definitions allow us to rewrite $E,F,G,H$ as\\

\begin{align}
E&=N\sum_{\ell} \ell p_\ell &F&=N\sum_\ell \ell q_\ell \label{eqnA:defs2.1}\\
G&=N^2\sum_\ell\sum_{\ell'} \min\{\ell,\ell'\}p_\ell p_{\ell'}&H&=N\sum_\ell \ell^2 q_\ell ,\label{eqnA:defs2.2}
\end{align}
so the expression for the variance of $\hat\theta$ can be written as:
\begin{align}
\begin{split}
N^{-1}\sigma_e^2\var(\hat\theta)^{-1}=&\sum_\ell \ell p_\ell-\sum_\ell \ell^2q_\ell -\sum_\ell \sum_{\ell'} \min\{\ell,\ell'\}p_\ell p_{\ell'}\label{eqnA:onjfun0}\\
&-\frac{T(\sum_\ell \ell q_\ell)^2+W(\sum_\ell \ell p_\ell)^2-2(\sum_\ell \ell p_\ell)(\sum_\ell \ell q_\ell)}{1-WT}.
\end{split}
\end{align}
 
It is convenient to express \eqref{eqnA:onjfun0} in terms of quadratic forms in $P$ and $Q$, the $S$-dimensional vectors with $\ell$th elements $p_\ell$ and $q_\ell$, respectively.  Using the fact that $\sum p_\ell=1$ and $\sum q_\ell=W$, \eqref{eqnA:onjfun0} can be written as 
\begin{equation}\label{eqnA:objfun1}
N^{-1}\sigma_e^2\var(\hat\theta)^{-1}=P^T\Xi P-Q^T\tilde\Lambda' P-\left (\frac{TQ^T\Delta' Q+WP^T\Delta' P-2Q^T\Delta' P}{1-WT}  \right ),
\end{equation}
with the matrices $\Xi,\Delta'$ and $\tilde\Lambda'$: $\Xi_{ij}=\tfrac{1}{2}\mid i-j \mid$, $\tilde\Lambda'=y'1^T$ and $\Delta'=z'z'^T$, with $z'^T=(1,2,\ldots,S)$ and $y'$ comprises the squares of the elements of $z'$, i.e. $y'^T=(1^2,2^2,\ldots,S^2)$.\\

Routine algebra shows that the expression \eqref{eqnA:objfun1} is unchanged if $\tilde\Lambda',\Delta'$ are replaced by $\tilde\Lambda,\Delta$, defined as $\tilde\Lambda=y1^T$ and $\Delta=zz^T$, where $z=z'-\tfrac{1}{2}(S+1)$ and $y$ comprises the squares of the elements of $z$.  This alternative expression, which is \eqref{eqn:objfun1}, makes certain symmetry properties of the derived designs more readily apparent.

\subsection*{Regression approximation}

\begin{figure}
\centering
\includegraphics[width=0.9\linewidth]{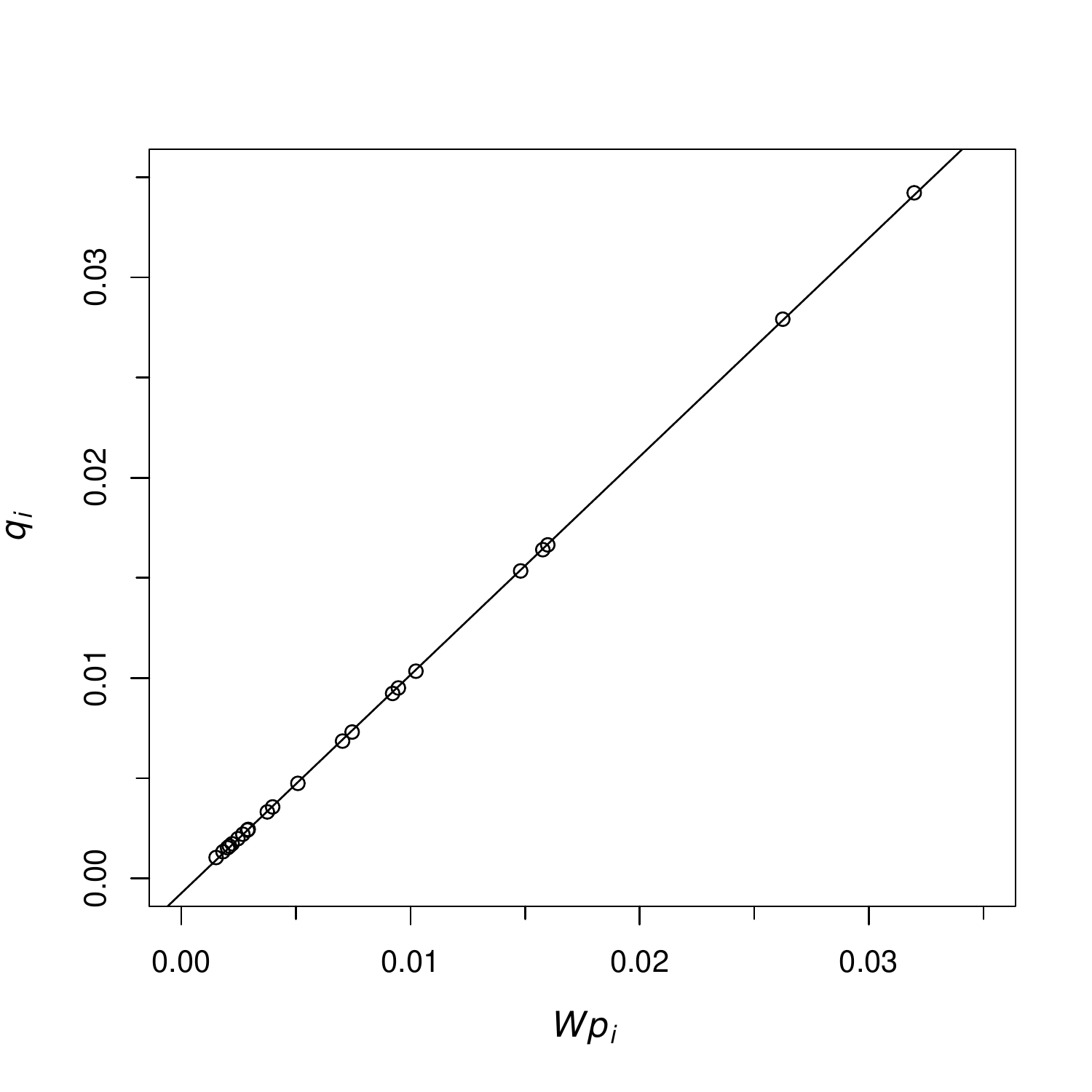}
\caption{The relationship between the $q_i$ and $Wp_i$ for the clusters used in \citep{sexpart}, with $\lambda=276$ and $T=5$. The line is the ordinary least squares regression.}\label{fig:fig1rev}
\end{figure}

Progress from \eqref{eqn:objfun1} is facilitated by noticing that the regression of $q_i$ on $WP_i$ has a very high correlations, which means very accurate and useful approximations for $Q$ are available.  Figure~\ref{fig:fig1rev} shows the regression of $q_i$ on $Wp_i$ for the 22 clusters from \citet{sexpart}. A very similar plot obtains for the RRT trial.  Both for these examples and for a variety of simulated examples, the close linear relation between $q_i$ and $Wp_i$ holds for a range of $N,\lambda$ and $T$ which includes those likely to be relevant to SWDs found in practice.
Eliminating $Q$ from \eqref{eqn:objfun1} using the regression approximation requires $Q$ to be replaced by $W(1-\beta)K+W\beta P+R$ (cf. \eqref{eqn:regvec}) and then any terms involving $R$ disregarded.  In the following $\Lambda=\tfrac{1}{2}(\tilde\Lambda+\tilde\Lambda^T)$.  The terms involving $Q$ in \eqref{eqn:objfun1} are
\begin{align*}
Q^T\tilde\Lambda P&=R^T\tilde\Lambda P+W(1-\beta)K^T\tilde\Lambda P+W\beta P^T\Lambda P\\
Q^T\Delta P&=R^T\Delta P + W(1-\beta)K^T\Delta P +W\beta P^T\Delta P\\
Q^T\Delta Q&=R^T(2W\beta\Delta P+2(W(1-\beta)\Delta K+\Delta R)+2\beta(1-\beta)W^2K^T\Delta P\\
&+ W^2(1-\beta)^2K^T\Delta K+W^2\beta^2P^T\Delta P.
\end{align*}
Substituting these into \eqref{eqn:objfun1} and dropping any term involving $R$ results in \eqref{eqn:mainapp}.
 
\subsubsection*{Proof that $\beta>1$}

The coefficient of $K^Ty$ in \eqref{eqn:mainapp} is $-W(1-\beta)$ and the sign of this quantity is important in determining the form of optimal and efficient designs.  It is positive because $W>0$ and $\beta>1$.  This second inequality is proved below.\\

The regression of $q_i$ on $Wp_i$ has slope $\beta$, and this exceeds 1 iff 
\[
\sum_{i=1}^C \frac{q_i}{W}(p_i-C^{-1})=\sum_{i=1}^C \frac{q_i}{W}p_1-C^{-1}>\sum_{i=1}^C (p_i-C^{-1})^2=\sum_{i=1}^C p_i^2-C^{-1}.
\]
So $\beta>1$ iff $\sum q_ip_i>W\sum p_i^2$, i.e. iff $\sum p_i\eta_i>0$, where $\eta_i=q_i-Wp_i$.  Now
\begin{align*}
\eta_ip_i&=q_ip_i-Wp_i^2\\
&=\frac{p_i^3}{\lambda'+p_iT}-\sum_{j}p^2_i\frac{p_j^2}{\lambda'+p_jT}&=\frac{p_i^3(1-p_i)}{\lambda'+p_iT}-\sum_{j\ne i}\frac{p^2_ip_j^2}{\lambda'+p_jT}\\
&=\frac{p_i^3\sum_{j\ne i} p_j}{\lambda'+p_iT}-\sum_{j\ne i}\frac{p^2_ip_j^2}{\lambda'+p_jT}&=\sum_{j\ne i}\left ( \frac{p_i^3p_j}{\lambda'+p_iT}-\frac{p_i^2p_j^2}{\lambda'+p_jT}\right )\\
&=\lambda'\sum_{j\ne i}\frac{p_i^2p_j(p_i-p_j)}{(\lambda'+p_jT)(\lambda'+p_iT)}&=\lambda'\sum_j\frac{p_i^2p_j(p_i-p_j)}{(\lambda'+p_jT)(\lambda'+p_iT)}\\
&=\lambda'\sum_j \psi_{ij}, \mathrm{say}.
\end{align*}
So $\sum_i \eta_ip_i=\lambda'\sum_i\sum_j \psi_{ij}$.  Note that 
\[
\psi_{ij}+\psi_{ji}=\frac{p_i^2p_j(p_i-p_j)+p_j^2p_i(p_j-p_i)}{(\lambda'+p_jT)(\lambda'+p_iT)}=\frac{p_ip_j(p_i-p_j)^2}{(\lambda'+p_jT)(\lambda'+p_iT)}>0,
\]  
thereby proving that $\beta>1$.
\section*{Appendix B: {\large designs which maximise $\mathscr{V}$}}

A useful calculation for both cluster-balanced and general allocations is to maximise \eqref{eqn:mainapp} taking $K$, and hence also $b$ and $a$, to be fixed.  To accommodate the constraint $1^TP=1$, define the Lagrangian
\[
\mathscr{L}=P^TAP +h_1 b z^TP+\nu 1^TP,
\]
with Lagrange multiplier $\nu$.  Then
\[
\frac{\partial \mathscr{L}}{\partial P}=0 \Rightarrow P=-\tfrac{1}{2}h_1bA^{-1}z-\tfrac{1}{2}\nu A^{-1}1
\]
Evaluating $\nu$ using $1^TP=1$ leads to the optimising $P=P_\mathrm{opt}$ being found to be
\[
P_\mathrm{opt}=\frac{A^{-1}1}{1^TA^{-1}1}+\frac{1}{2}\frac{h_1b}{1^TA^{-1}1}(1^TA^{-1}zA^{-1}1-1^TA^{-1}1A^{-1}z).
\]
An explicit formula requires expressions for $A^{-1}1$ and $A^{-1}z$: these are available and will be evaluated below

\subsection*{Evaluating $A^{-1}1$ and $A^{-1}z$}
Define $e,f$ as $S$-dimensional vectors that are zero except for the first and last elements, which are $e_1=f_1=1$ and $e_S=-f_S=1$.  The key results needed are
\begin{align*}
\Xi 1&=\tfrac{1}{2}y+\tfrac{1}{8}(S^2-1)1&\Xi e&=\tfrac{1}{2}(S-1)1&\Xi f&=z\\
\Lambda 1&=\tfrac{1}{2}Sy+\tfrac{1}{24}S(S^2-1)1&\Lambda e&=y+\tfrac{1}{4}(S-1)^21&\Lambda f&=0\\
\Delta 1&=0&\Delta e&=0&\Delta f&=-(S-1)z
\end{align*}
From these results it is clear that $A(a1+be+cf)$ is a linear combination of $1,z,y$.  Setting $A(a1+be+cf)=1$ or $z$ and solving for $a,b,c$ provides $A^{-1}1$ and $A^{-1}z$, which are
\begin{align*}
A^{-1}1&=\frac{12W\beta 1+6(1-W\beta S)e}{(S-1)(3-3(S-1)W\beta+S(S-2)W^2\beta^2)},\\
A^{-1}z&=\frac{1}{1-\gamma W(S-1)}f.
\end{align*} 
From these the expression in \eqref{eqn:fullopt} can be obtained.\\

\subsection*{Expression for $h_3$ and its sign}

Substituting $P=P_\mathrm{opt}$ into $\mathscr{V}(P,K)$ gives $\mathscr{V}_\mathrm{opt}=\frac{1}{1^TA^{-1}1}-h_3b^2-W(1-\beta)a$, where
 \[
h_3=h_2-\frac{h_1^2(S-1)}{4(1-\gamma W(S-1))}.
\]
The role of $b$ in maximising $\mathscr{V}$ depends on the sign of $h_3$, which is shown below to be positive.  First, it is useful to show that the denominator of the second term of $h_3$ is positive.

\subsubsection*{Showing that $1-\gamma W(S-1)>0$}
This follows because $\gamma W(S-1)<1$ provided $2\beta-1-\beta^2WT)W(S-1)<1-WT$,  The left hand side (LHS) is a quadratic in $\beta$ which is maximised when $\beta=1/(WT)$.  So the LHS has maximum 
\[
(\frac{2}{WT}-1-\frac{1}{WT})W(S-1)=(1-WT)\frac{W(S-1)}{WT}
\]
and this is less than $1-WT$ because $(S-1)<T=S+1$. 

\subsubsection*{Showing that $h_3>0$}
This will follow if
\[
\frac{h_1^2(S-1)}{4(1-\gamma W(S-1))h_2}=\frac{(1-W\beta T)^2(S-1)}{T(1-WT)(1-\gamma W(S-1))}<1.
\]
This follows provided
{\small
\begin{align*}
(1-W\beta T)^2(S-1) &<T\left [(1-WT-(2\beta-1-\beta^2WT)W(S-1)\right ]\\
\shortintertext{i.e. provided}
(S-1)-2T(S-1)W\beta+(S-1)T^2W^2\beta^2&<T[1-WT+W(S-1)]\\
&-2T(S-1)W\beta+(S-1)T^2W^2\beta^2\\
\shortintertext{and recalling that $S=T-1$, this inequality amounts to}
\end{align*}
\[
T-2<T[1-WT+W(T-2)]=T-2WT
\]
}
and this is true because $WT<1$.\\
\section*{Appendix C: {\large effect of $\lambda$ on designs and approximations for $W$ and $W\beta$}}

\subsection*{How $\mathscr{V}$ varies with $\lambda$}

The values of $\mathscr{V}$ for the designs $D1$ and $D2$ described in Section~\ref{sect:numres} are shown in Table~\ref{A:tab1} for a range of $\lambda$.  It can be seen in Table~\ref{A:tab1} that the difference between $\mathscr{V}_{D1}$ and $\mathscr{V}_{D2}$ is smallest for small and large $\lambda$.  The difference between the designs is only due to the different numbers of clusters on each sequence (so $a=1.25$ for $D1$ and $a=1.75$ for $D2$), as the number of individuals on each sequence is the same in the two designs. The size of the difference between the designs also depends on the coefficient of $a$, which is $-W(1-\beta)$, and is also shown in Table~\ref{A:tab1}.  The pattern of differences arises because when $\lambda$ is large the observations are essentially independent and their division into clusters is immaterial.  When $\lambda$ is small the term $\alpha_i$ has a relatively high variance, so the form of the estimator for $\theta$ will lessen its dependence on between-cluster information and, when $\lambda=0$, $\alpha_i$ is eliminated from $\hat\theta$ \citep[see][]{matthewsforbes}, which is why the division into clusters is also unimportant when $\lambda$ is small.
\begin{table}[h]
\begin{center}
\caption{Values of $\mathscr{V_{D1}}$ $\mathscr{V_{D2}}$ for various $\lambda$: the range of $\lambda$ corresponds to the ICC going from 0.667 to 0.0002. Note $a=1.25$ for $D1$ and $a=1.75$ for $D2$}
\label{A:tab1}
\begin{tabular}{c|ccccc}
&\multicolumn{5}{c}{$\lambda$}\\
$\mathscr{V}$ & 0.5 & 5 & 50 & 500 &5000\\
\hline
$\mathscr{V}_{D1}$& 0.380 & 0.394& 0.486& 0.643&0.688\\
$\mathscr{V}_{D2}$& 0.380 & 0.399 & 0.508 &0.653 & 0.690\\
$W(1-\beta)$&-0.001&-0.012&-0.044&-0.020&-0.0003
\end{tabular}
\end{center}
\end{table}

\subsection*{Approximations to $W$ and $W\beta$}

If the $N_i$ are not known {\it a priori} and only values for their mean, $M$ and coefficient of variation $CV$ are to hand, then the following approximations may prove useful.

\subsubsection*{Approximating $W$}

If we assume that the $N_i$ are a random selection from some population, then $\expect(W)$ can be approximated by $C\expect(q_i)$.  Recall that $q_i=f(p_i)$, where $f(x)=x^2/(k+xT)$ and $k=\lambda/N$.  As the $p_i$ sum to 1, $\expect(p_i)=C^{-1}$, and expanding $f(\cdot)$ about its mean gives
\[
\expect(W)\approx C\left (f(C^{-1})+\tfrac{1}{2}f''(C^{-1})\var(p_i)\right ).
\]  
As $f''(x)=2k^2(k+xT)^{-3}$, this can be written as
\begin{align*}
\expect(W)&\approx\frac{C^{-1}}{k+C^{-1}T}+C\frac{k^2}{(k+C^{-1}T)^3}\var(p_i)\\
&=\frac{N}{C\lambda+NT}+\frac{\lambda^2NC^2}{(C\lambda+NT)^3}CV^2\\
\shortintertext{and, as $M=NC^{-1}$,}
&=\frac{M}{\lambda+MT}+\frac{\lambda^2M}{(\lambda+MT)^3}CV^2.
\end{align*}
For the EPT trial the mean cluster size is 495.23, $CV=0.9975$ and $W=0.1816$ ($\lambda=276$) and $W=0.1984$ ($\lambda=20$).  Using just the first term of the above gives $W\approx 0.1799$ ($\lambda=276$) and $W\approx 0.1984$ ($\lambda=20$).  With the second term included $W\approx 0.1817$ ($\lambda=276$) and $W\approx 0.1984$ ($\lambda=20$). For the RRT study the mean cluster size is 4.667, $CV=0.3499$ and $W=0.1710$ ($\lambda=9$) and $W=0.1276$ ($\lambda=19$). Using the first term  gives $W\approx 0.1687$ ($\lambda=9$) and $W\approx 0.1239$ ($\lambda=19$).  With the second term included $W\approx 0.1709$ ($\lambda=9$) and $W\approx 0.1278$ ($\lambda=19$).

\subsubsection*{Approximating $W\beta$}

A slightly different approach is used here.  From the definition of $\beta$ we have $W\beta=\cov(q_i,p_i)/\var(p_i)$.  As $q_i$ and $p_i$ are very highly correlated, then by assuming that this correlation very close to is 1,  $\cov(p_i,q_i)\approx\sqrt{\var(q_i)\var(p_i)}$, so $W\beta\approx\sqrt{\var(q_i)/\var(p_i)}$.  As $q_i=f(p_i)$, the delta method will approximate this ratio by $f'(C^{-1})$. Now
\[
f'(x)=\frac{2kx+x^2T}{(k+xT)^2} \Rightarrow Tf'(x)=1-\frac{k^2}{(k+xT)^2}.
\]
Hence
\[
f'(C^{-1})=\frac{1}{T}\left ( 1-\frac{\lambda^2}{(\lambda+C^{-1}NT)^2}\right )=\frac{1}{T}\left ( 1-\frac{\lambda^2}{(\lambda+MT)^2}\right ).
\]
A range of simulations showed that $W$ and $\beta$ have a noticeable negative correlation, leading to $W\beta$ being less variable than $W$.
For the EPT trial $W\beta=0.1980$ ($\lambda=276$) and $W\beta=0.199985$ ($\lambda=20$): the above approximation gives $W\beta\approx0.1980$ ($\lambda=276$) and $W\beta\approx0.199987$ ($\lambda=20$). In the EPT trial $W\beta=0.2162$ ($\lambda=9$) and $W\beta=0.1758$ ($\lambda=19$) and the approximation gives $W\beta\approx0.2235$ ($\lambda=9$) and $W\beta\approx0.1864$ ($\lambda=19$).

\section*{Appendix D: {\large the closed-cohort design}}

The model for the closed-cohort design, namely
\begin{equation*}\label{eqn:modelccA}
Y_{ijk}=\beta_j + X_{ij}\theta + \alpha_i +\zeta_{ik}+\epsilon_{ijk},
\end{equation*}
gives rise to a vector of responses on an individual of $(Y_{i1k},Y_{i2k},\ldots,Y_{iTk})^T$, $k=1,\ldots,N_i$ and the mean of these for cluster $i$ is $Y_i$, which has variance $V_i=(\tau^2+N_i^{-1}\omega^2)J_T+N_i^{-1}\iv I_T$, with inverse
\[
V_i^{-1}=\frac{N_i}{\iv}(I_T-\tilde w_i J_T); \;\; \tilde w_i=\frac{N_i+\mu}{\lambda +(N_i+\mu)T},
\]
where $\mu=\omega^2/\tau^2$.  It follows that $\sum V_i^{-1}=N\sigma_e^{-2}(I_T-\tilde W J_T)$, where 
\[
\tilde W=\sum_{i=1}^C \frac{N_i}{N}\tilde w_i.
\]
Also note that 
\[
(\sum V_i^{-1})^{-1}= \frac{\iv}{N}(I_T+\frac{\tilde W}{1-\tilde WT}J_T).
\]

It is also the case that $V_i^{-1}D_i=\sigma_e^{-2}N_i(D_i-\tilde w_ir_i1_T)$.\\

These expressions are all identical to those found for the cross-sectional case, provided $w_i$ is replaced with $\tilde w_i$.  The calculations which led to the different forms of $\var(\hat\theta)$ for the cross-sectional case are the same for the cohort case if $W$ is replaced with $\tilde W$ and $q_i$($q_\ell)$ are replaced with $\tilde q_i$ ($\tilde q_\ell$), where
\[
\tilde q_i=\frac{N_i}{N}\tilde w_i=\frac{N_i(N_i+\mu)}{N[\lambda+(N_i+\mu)T]}=\frac{p_i(p_i+\mu')}{\lambda'+(p_i+\mu')T},
\]
where $\mu'=\mu/N$. Provided that the values of $\lambda',\mu'$ are such that the correlation between $\tilde q_i$ and $p_i$ is high, then the methods developed for the cross-sectional case also apply here.

\end{document}